\documentstyle[12pt,aps,prd,preprint]{revtex}
\begin{document}
\title{Radiative Tail of Realistic Rotating Gravitational Collapse}
\author{Shahar Hod}
\address{The Racah Institute for Physics, The
Hebrew University, Jerusalem 91904, Israel}
\date{\today}
\maketitle

\begin{abstract}
  An astrophysically realistic model of wave dynamics in black-hole
  spacetimes must involve a {\it non}-spherical background geometry
  with {\it angular momentum}.  We consider the evolution of {\it
    gravitational} (and electromagnetic) perturbations in {\it
    rotating} Kerr spacetimes.  We show that a rotating Kerr black
  hole becomes ``bald'' {\it slower} than the corresponding
  spherically-symmetric Schwarzschild black hole. Moreover, our
  results {\it turn over} the traditional belief (which has been widely accepted
  during the last three decades) that the late-time tail of
  gravitational collapse is universal.  In particular, we show that
  different fields have {\it different} decaying rates.  Our results
  are also of importance both to the study of the no-hair
  conjecture and the mass-inflation scenario (stability of Cauchy
  horizons).

\end{abstract}
\bigskip

The radiative tail of gravitational collapse decays with time leaving
behind a Kerr-Newman black hole characterized solely 
by the black-hole mass, charge, and angular-momentum. This is the
essence of the {\it no-hair} conjecture, introduced by Wheeler in the
early 1970s \cite{Whee}. 

Price \cite{Price} was the first to analyze the mechanism by which the
spacetime outside a (nearly {\it spherical}) star divests itself of
all radiative multipole moments, and leaves behind a Schwarzschild
black hole; it was demonstrated that all radiative perturbations decay
asymptotically as an inverse power of time, the power indices equal 
$2l+3$ (in absolute value), where $l$ is the multipole order of the
perturbation. This late-time decay of radiative fields is often
referred to as their ``power-law tail''.  Physically, these inverse
power-law tails are associated with the backscattering of waves off
the effective curvature potential at asymptotically far regions
\cite{Thorne,Price}.

The analysis of Price has been extended by many authors. 
We shall not attempt to review the numerous works 
which address the problem of the late-time evolution of
gravitational collapse. For a partial list of references, see e.g.,
\cite{Hod2}. These earlier analyses were restricted, 
however, to {\it spherically} symmetric backgrounds. 
It is well-known that realistic stellar objects generally rotate about their
axis, and are therefore not spherical. Thus, 
the nature of the physical process of stellar core collapse to
form a black hole is essentially {\it non-}spheric. An astrophysically
realistic model must therefore take into account the angular momentum of 
the background geometry.

The corresponding problem of wave dynamics in 
realistic {\it rotating} Kerr spacetimes is much more complicated due to
the lack of spherical symmetry. A first progress has
been achieved only recently \cite{Krivan1,Krivan2,Ori,Barack2}. 
Detailed analyses for the simplified toy-model of a test {\it scalar} field in the
Kerr background have been given recently in \cite{Hod1,BarOr}. 

Obviously, the most interesting situation from a physical point of
view is the dynamics of {\it gravitational} perturbations
in  {\it rotating} Kerr spacetimes. This is the subject of this
Letter, in which we present our main results for this fascinating 
problem. Full details of the
analysis are given elsewhere \cite{Hod2}.

The dynamics of massless perturbations outside a realistic rotating
Kerr black hole is governed by Teukolsky's 
master equation \cite{Teukolsky1,Teukolsky2}:

\begin{eqnarray}\label{Eq1}
&& \left[{{(r^{2}+a^{2})^{2}} \over {\Delta}} 
-a^{2}\sin^{2}\theta \right]
{{\partial ^2 \psi} \over {\partial t^2}} 
+{{4Mar} \over {\Delta}} 
{{\partial ^2 \psi} \over {\partial t \partial \varphi}}
+ \left[{{a^{2}} \over \Delta} -{1 
\over {\sin^{2} \theta}} \right] {{\partial ^2 \psi} 
\over {\partial \varphi ^2}} \nonumber \\
&& -\Delta^{-s} {\partial \over {\partial r}} 
\left( \Delta^{s+1} {{\partial \psi} 
\over {\partial r}} \right) -{1 \over {\sin\theta}} 
{{\partial} \over {\partial
    \theta}} \left( \sin \theta{{\partial \psi} \over {\partial
    \theta}} \right) -2s \left[{{a(r-m)} \over \Delta} 
+{{i\cos \theta} \over {\sin^{2} \theta
}} \right] {{\partial \psi} \over {\partial
    \varphi}} \nonumber \\
&& -2s \left[{{M(r^{2}-a^{2})} \over \Delta} -r -ia\cos 
\theta \right] {{\partial \psi} \over
 {\partial t}} +(s^{2}\cot^{2} \theta -s) \psi =0\ \ ,
\end{eqnarray}
where $M$ and $a$ are the mass and angular-momentum per unit-mass of
the black hole, and $\Delta=r^{2} -2Mr+a^{2}$. (We use gravitational units in which $G=c=1$).
The parameter $s$ is called the spin-weight of the field. For
gravitational perturbations $s= \pm 2$ (for electromagnetic
perturbations $s= \pm 1$).
The field quantities $\psi$ which satisfy Teukolsky's equation 
are given in \cite{Teukolsky2}.

Resolving the field in the form $\psi= \Delta^{-s/2} (r^{2}+a^{2})^{-1/2} 
\sum\limits_{m= -\infty}^{\infty} 
\Psi^{m} e^{im \varphi}$ (where $m$ is the azimuthal number), 
one obtains a wave-equation for each value of $m$

\begin{equation}\label{Eq2}
D \Psi \equiv \left [ B_{1} {{\partial ^2} \over {\partial t^2}} +
B_{2} {{\partial} \over
{\partial t}} -
{{\partial ^2} \over {\partial y^2}}+ B_{3} - 
{{\Delta} \over {(r^{2}+a^{2})^2}} {1 \over {\sin\theta}} 
{{\partial} \over {\partial
    \theta}} \left( \sin \theta{{\partial} \over {\partial
    \theta}} \right) \right ] \Psi=0\  ,
\end{equation}
where the tortoise radial coordinate $y$ is defined 
by $dy=[(r^2+a^2)/\Delta] dr$. (We suppress the index $m$).
The coefficients $B_i=B_{i}(r,\theta)$ are given by

\begin{equation}\label{Eq3}
B_{1}(r,\theta)=1-{{\Delta a^{2}\sin^{2}\theta} \over {(r^{2}+a^{2})^{2}}}\ \ ,
\end{equation}
and

\begin{equation}\label{Eq4}
B_{2}(r,\theta)=\Bigg \{ {{4iMmar} \over \Delta} -2s \left[
  {{M(r^{2}-a^{2})} 
\over \Delta} -r -ia\cos \theta \right] \Bigg \} {\Delta \over
{(r^2+a^2)^2}}\ \ .
\end{equation}
[The explicit expression of $B_3(r,\theta)$ is not important for the analysis].

The time-evolution of a wave-field described by Eq. (\ref{Eq2}) is given by

\begin{eqnarray}\label{Eq5}
\Psi (z,t) &=& 2\pi \int \int_{0}^{\pi} \Bigg\{ B_{1}(z') 
\Big [ G(z,z';t) \Psi _t(z',0)+G_t(z,z';t) 
\Psi (z',0) \Big] +  \nonumber \\ 
&& B_{2}(z') G(z,z';t) \Psi (z',0) \Bigg\} \sin\theta' d\theta'dy'\ \ ,
\end{eqnarray}
for $t>0$, where $z$ stands for $(y,\theta)$. 
The (retarded) Green's function $G(z,z';t)$ is defined by 
$DG(z,z';t)=\delta (t) \delta(y-y') {{\delta(\theta - \theta')}/ {2\pi
    \sin\theta}}$ with $G=0$ for $t<0$. 
We express the Green's function in terms of the the Fourier transform
$\tilde G_{l}(y,y';w)$

\begin{equation}\label{Eq6}
G(z,z';t)={1 \over {(2 \pi)^{2}}} \sum\limits_{l=l_{0}}^{\infty} 
\int_{- \infty +ic}^{\infty +ic}
\tilde G_{l}(y,y';w)_{s}S_{l}^{m}(\theta,aw)_{s}S_{l}^{m}(\theta',aw)
e^{-iwt} dw\  ,
\end{equation}
where $c$ is some positive constant and $l_{0}=max(|m|,|s|)$. 
The functions $_sS_{l}^{m}(\theta,aw)$ are the spin-weighted
spheroidal harmonics
which are solutions to the angular equation \cite{Teukolsky2}

\begin{eqnarray}\label{Eq7}
&& {1 \over {\sin\theta}} 
{{\partial} \over {\partial
    \theta}} \left( \sin \theta{{\partial} \over {\partial
    \theta}} \right) + \nonumber \\
&&  \left(a^{2}w^{2}\cos^{2}\theta-{{m^{2}} \over
  {\sin^{2}\theta}} -2aws\cos \theta - {{2ms\cos\theta} \over
  {\sin^{2}\theta}} - s^{2}\cot^{2}\theta + s +{_{s}A_{l}^{m}} \right) 
{ _{s}S_{l}^{m}}  =0  \  .
\end{eqnarray}

The Fourier transform is analytic in the upper half $w$-plane and it satisfies
the equation \cite{Teukolsky2}

\begin{eqnarray}\label{Eq8}
 \tilde D(w) \tilde G_l  &\equiv&  \Bigg \{{{d^2} \over {dy^2}} + \left[ {{K^{2} -2is(r-M)K+
    \Delta(4irws-\lambda)} \over {(r^{2}+a^{2})^{2}}}-H^{2}-{{dH} \over
    {dy}} \right] \Bigg\} \tilde G_{l}(y,y';w) \nonumber \\
& =& \delta(y-y')  \  ,
\end{eqnarray}
where $K=(r^{2}+a^{2})w-am$, $\lambda=A+a^{2}w^{2}-2amw$, and 
$H=s(r-M)/(r^{2}+a^{2})+r\Delta/(r^{2}+a^{2})^{2}$.

Define two auxiliary functions $\tilde \Psi_1$ and $\tilde \Psi_2$ as 
solutions to the homogeneous equation
$\tilde D(w)\tilde \Psi_1$=$\tilde D(w)\tilde \Psi_2=0$, with the physical boundary
conditions of purely ingoing waves crossing the event horizon, and
purely outgoing waves at spatial infinity, respectively. In terms of
$\tilde \Psi_1$ and $\tilde \Psi_2$, and henceforth assuming $y'<y$, 
$\tilde G_{l}(y,y';w) =-\tilde \Psi_1(y',w) \tilde \Psi_2(y,w)/W(w)$,
where we have used the Wronskian 
relation $W(w)=W(\tilde \Psi_1, \tilde \Psi_2)= \tilde \Psi_1 \tilde \Psi_{2,y} - 
\tilde \Psi_2 \tilde \Psi_{1,y}$.

It is well known that the late-time behaviour of massless perturbations 
fields is determined by the backscattering from asymptotically {\it far}
regions \cite{Thorne,Price}. Thus, the late-time behaviour is dominated by the
{\it low}-frequencies contribution to the Green's function, for only low
frequencies will be backscattered by the small effective curvature 
potential (at $r \gg M$). Therefore, a {\it small}-$w$ approximation
[or equivalently, a large-$r$ approximation of Eq. (\ref{Eq8})] is sufficient in order
to study the asymptotic {\it late-time} behaviour of the fields 
\cite{Andersson}. With this approximation, the two basic solutions required in order to build 
the Fourier transform 
are $\tilde \Psi_1 =r^{l +1} e^{iwr} M(l+s+1-2iwM ,2l +2, 
-2iwr)$ and $\tilde \Psi_2 =r^{l +1} e^{iwr} U(l+s+1-2iwM ,2l +2, 
-2iwr)$, where $M(a,b,z)$ and $U(a,b,z)$ are 
the two standard solutions to the confluent hypergeometric equation 
\cite{Abram}. Then $W(\tilde \Psi_1,\tilde \Psi_2)=i (-1)^{l+1} (2l+1)! (2w)^{-(2l +1)}/(l+s)!$. 

In order to calculate $G(z,z';t)$ using Eq. (\ref{Eq6}), one may 
close the contour of integration into the lower half of the
complex frequency plane. Then, one identifies three distinct contributions to $G(z,z';t)$
\cite{Leaver} : Prompt contribution, quasinormal modes, and tail
contribution. The late-time tail is associated with the existence of a
branch cut (in $\tilde \Psi_2$) in the complex frequency plane \cite{Leaver} (usually
placed along the negative imaginary $w-$ axis). A little arithmetic leads to \cite{Hod3}

\begin{eqnarray}\label{Eq9}
\tilde G_l^C(y,y';w)&=& \Bigg[{{\tilde \Psi_2(y,we^{2 \pi i})} \over {W(we^{2 \pi i})}} -
{{\tilde \Psi_2(y,w)} \over {W(w)}} \bigg]\tilde \Psi_1(y',w)
\nonumber \\
& =&{{(-1)^{l-s} 4 \pi Mw (l-s)!} \over {(2l+1)! }}
{{\tilde \Psi_1(y,w)\tilde \Psi_1(y',w)} \over {W(w)}}\  .
\end{eqnarray}
Taking cognizance of
Eq. (\ref{Eq6}) we obtain

\begin{eqnarray}\label{Eq10}
G^C(z,z';t)& =&  \sum\limits_{l=l_0}^{\infty} 
{{i M (-1)^s 2^{2l+1} (l+s)! (l-s)!} \over {\pi [(2l+1)!]^2}} \nonumber \\
&& \int_{0}^{-i \infty} \tilde \Psi_1(y,w) 
\tilde \Psi_1(y',w) {_{s}S_{l}(\theta,aw)} {_{s}S_{l}(\theta',aw)}
w^{2l+2} e^{-iwt} dw\  .
\end{eqnarray}

The angular equation (\ref{Eq7}) is amenable to a perturbation treatment for small
$aw$; we write it in the form 
$(L^{0}+L^{1}){_sS_{l}^{m}}=-{_sA_{l}^{m}}{_sS_{l}^{m}}$, where
$L^{1}(\theta,aw) \equiv (aw)^2 \cos^{2}\theta -2aws\cos \theta$ 
[and $L^{0}(\theta)$ is the $w$-independent part of Eq. ({\ref{Eq7})], and we use the 
spin-weighted spherical functions $_sY_{l}^{m}$ as a
representation. They satisfy $L^{0}{_sY_{l}}=-{_sA_{l}^{(0)}}{_sY_{l}}$ 
with $_sA_{l}^{(0)}=(l-s)(l+s+1)$. 
For small $aw$ a standard perturbation 
theory (see, for example, \cite{Schiff}) yields \cite{Hod2}

\begin{equation}\label{Eq11}
_sS_{l}(\theta,aw)= \sum\limits_{k=l_0}^{\infty} C_{lk}(aw)^{|l-k|} {_sY_{k}(\theta)}\  ,
\end{equation}
where, to leading order in $aw$, the 
coefficients $C_{lk}(aw)$ are $w-${\it independent} \cite{Hod2}. 
Equation (\ref{Eq11}) implies that the black-hole rotation {\it mixes}
different spin-weighted spherical harmonics.

The time-evolution of the fields is given by Eq. (\ref{Eq5}).
Therefore, in order to elucidate the coupling between different modes we should
evaluate the integrals  
$\langle slm|skm \rangle$, $\langle slm|\sin^2\theta|skm \rangle$, and
$\langle slm|\cos\theta|skm \rangle$, where 
$\langle slm|F(\theta)|skm \rangle \equiv \int {_sY_l^{m*}}
F(\theta){_sY_k^m} d \Omega$ 
[see Eqs. (\ref{Eq3}) and (\ref{Eq4}) 
for the definition of the $B_i(r,\theta)$ coefficients]. 
The spin-weighted spherical harmonics are related to the rotation
matrix elements of quantum mechanics \cite{CamMor}. Hence, standard
formulae are available for integrating the product of three such
functions (these are given in terms of the Clebsch-Gordan
coefficients \cite{Hod2}). In particular, the integral 
$\langle sl0|\sin^2\theta|sk0 \rangle$ vanishes 
unless $l=k,k \pm 2$, while the integral $\langle
sl0|\cos\theta|sk0 \rangle$ vanishes unless $l=k \pm 1$. 
For non-axially symmetric $(m \neq 0)$ modes, 
$\langle slm|\sin^2\theta|skm \rangle \neq 0$ for $l=k, k \pm 1, k \pm 2$, and 
$\langle slm|\cos\theta|skm \rangle \neq 0$ for $l=k, k \pm 1$ (all
other matrix elements vanish).

{\it Asymptotic behaviour at timelike infinity.} 
As already explained, the late-time behaviour of the fields should follow from the
{\it low}-frequency contribution to the Green's function. Actually, it is easy 
to verify that the effective contribution to the integral in 
Eq. (\ref{Eq10}) should come from $|w|$=$O({1/t})$. 
Thus, in order to obtain the asymptotic behaviour of the fields 
at {\it timelike infinity} (where $y,y' \ll t$) 
we may use the $|w|r \ll 1$ asymptotic limit of $\tilde \Psi_1(r,w)$, which is
given by $\tilde \Psi_1(r,w) \simeq r^{l +1}$ \cite{Abram}. 

Substituting this in Eq. (\ref{Eq10}), and 
using the representation Eq. (\ref{Eq11}) for the spin-weighted 
spheroidal wave functions $_sS_{l}$ [togather with the above cited
properties of the integrals $\langle slm|\sin^2 \theta|skm \rangle$ and 
$\langle slm|\cos\theta|skm \rangle$], we find that 
the asymptotic late-time behaviour of 
the $l$ mode (where $l \geq l_0$)
is {\it dominated} by the following effective Green's function:

\begin{equation}\label{Eq12}
G^C_l(z,z';t) =  M F_1 (yy')^{l_0+1} {_sY_{l}}(\theta) {_sY^{*}_{l_0}}(\theta')
a^{l-l_0} t^{-(l+l_0+3)}\  ,
\end{equation}
where $F_1=F_1(l,l_0,m,s)=(-1)^{(l+l_0+2s+2)/2} 2^{2l_0+1} (l+l_0+2)! (l_0+s)! (l_0-s)!C_{{l_0}l}/\pi [(2l_0+1)!]^2$.
We emphasize that the power indices $l+l_0+3$ found here for 
{\it rotating} Kerr spacetimes are {\it smaller} than the
corresponding power indices (the well known $2l+3$)
in {\it spherically} symmetric Schwarzschild 
spacetimes. (There is an equality only for the $l=l_0$ mode). 
This implies a {\it slower} decay of perturbations in rotating Kerr
spacetimes. 

{\it Asymptotic behaviour at future null infinity.} 
It is easy to verify that for this case
the effective frequencies contributing to the integral in
Eq. (\ref{Eq10})  are of order $O({1/u})$.
Thus, for $y-y' \ll t \ll 2y-y'$ one may use the
$|w|y' \ll 1$ limit for $\tilde \Psi_1(y',w)$ and 
the $M \ll |w|^{-1} \ll y$ ($Imw < 0$) asymptotic limit of $\tilde \Psi_1(y,w)$, which is given
by $\tilde \Psi_1(y,w) \simeq e^{iwy} (2l+1)! 
e^{-i \pi (l+s+1)/2}(2w)^{-(l+s+1)}y^{-s}/(l-s)!$ \cite{Abram}.

Substituting this in Eq. (\ref{Eq10}), and using the 
representation Eq. (\ref{Eq11}) for the spin-weighted 
spheroidal wave functions, we find that the 
behaviour of the $l$ mode (where $l \geq l_0$) at 
the asymptotic region of null infinity $scri_+$ 
is dominated by the following effective Green's function:

\begin{equation}\label{Eq13}
G^C_l(z,z';t) =  \sum\limits_{k=l_0}^{l} 
M F_2 y'^{k+1}v^{-s} {_sY_l}(\theta) {_sY^{*}_k}(\theta')
a^{l-k}  u^{-(l-s+2)}\  ,
\end{equation}
where $F_2=F_2(l,k,m,s)=(-1)^{(l+k+2s+2)/2} 2^{k} (k+s)! (l-s+1)!C_{kl}/\pi (2k+1)!$.

{\it Asymptotic behaviour at the black-hole outer horizon.} 
The asymptotic solution to the homogeneous equation 
$\tilde D(w)\tilde \Psi_1(y,w)=0$ 
at the black-hole outer horizon $H_+$
($y \to -\infty$) is 
$\tilde \Psi_1(y,w)=C(w) \Delta^{-s/2} e^{-i(w-mw_+)y}$ \cite{Teukolsky2}, 
where $w_+=a/2Mr_+$ [$r_+=M+(M^2-a^2)^{1/2}$ is the location of the
black-hole outer horizon]. In addition, 
we use $\tilde \Psi_1(y',w) \simeq y'^{l+1}$. Regularity of the
solution requires $C$ to be an analytic function of $w$. We thus
expand $C(w)=C_0+C_1w+\cdots$ for small $w$ (as already explained, 
the late-time behaviour of the field is dominated by the {\it low}-frequency
contribution to the Green's function). 
Substituting this in Eq. (\ref{Eq10}), and 
using the representation Eq. (\ref{Eq11}) for the spin-weighted 
spheroidal wave functions, we find that the asymptotic 
behaviour of the $l$ mode (where $l \geq l_0$) at 
the black-hole outer horizon $H_+$ 
is dominated by the following effective Green's function:

\begin{equation}\label{Eq14}
G^C_l(z,z';t) =  
{_s\Gamma_l} M F_1 \Delta^{-s/2} y'^{l_0+1}
{_sY_{l}}(\theta) {_sY^{*}_{l_0}}(\theta')
a^{l-l_0} e^{imw_+y} v^{-(l+l_0+3+b)}\  ,
\end{equation}
where $_s\Gamma_l$ are constants, and $b=0$ {\it generically}, 
except for the unique case $m=0$ with $s>0$, in which $b=1$ \cite{BarOr2}.

{\it Pure initial pulse}. So far we have assumed that the initial pulse consists of all the
allowed ($l \geq l_0$) modes. If, on the other hand, the initial
angular distribution is characterized by a {\it pure} spin-weighted
spherical harmonic function $_sY_{l^*}^m$, then the 
asymptotic late-time tails are dominated 
by modes which, in general,
have an angular distribution {\it different} from the original one 
(a full analysis of this case is given in\cite{Hod2}). 
We find that the field's behaviour at the asymptotic regions of timelike
infinity $i_+$ and at the black-hole outer horizon $H_+$ is 
dominated again by the lowest allowed mode (i.e., $l=l_0$). 
The damping exponents are (in absolute value) $l^* + l_0+3-q$ and $l^* + l_0+3-q+b$,
respectively, where $q=min(l^*-l_0,2)$. 

On the other hand, the behaviour
of gravitational (and electromagnetic) perturbations at the
asymptotic region of null infinity $scri_+$ is dominated by the
$l=l_0$ mode if $l_0 \leq l^* \leq l_0+2$ and by the 
$l_0 \leq l \leq l^*-2$ modes otherwise \cite{Hod2}. 
The corresponding damping exponents 
are  $l_0-s+2$ and $l^*-s$, respectively.

{\it Summary and physical implications}. We have analyzed the 
dynamics of {\it gravitational} 
(physically, the most interesting case) 
and electromagnetic fields in realistic {\it rotating} black-hole
spacetimes. The main results and their physical implications are as follows:

(1) We have shown that the late-time evolution of 
realistic rotating gravitational collapse is characterized by 
inverse power-law decaying tails at the three asymptotic
regions: timelike infinity $i_{+}$, future null infinity $scri _{+}$, 
and the black-hole outer-horizon $H_{+}$ (where the power-law
behaviour is multiplied by an oscillatory term, caused by the dragging
of reference frames at the event horizon). The relaxation of the
fields is in accordance with the {\it no-hair} conjecture
\cite{Whee}. This Letter reveals the {\it dynamical} physical
mechanism behind this conjecture in the context of rotating
gravitational collapse.

(2) The {\it unique} and important feature of {\it rotating}
gravitational collapse is the active {\it coupling} between modes 
of {\it different} $l$ (but the same $m$). Physically,
this phenomena is caused by the dragging of reference frames, due to
the black-hole (or star's) rotation (this phenomena is absent in the
non-rotating $a=0$ case). As a consequence, the late-time evolution of
realistic rotating gravitational collapse has an angular distribution 
which is generically different from the original angular distribution
(in the initial pulse). 

(3) The power indices at a fixed radius are found to be
$l+l_0+3$. These damping exponents are generically {\it smaller} than the
corresponding power indices in spherically symmetric spacetimes. This
implies a {\it slower} decay of perturbations in rotating Kerr
spacetimes. Stated in a more pictorial
way, a rotating Kerr black hole generically becomes ``bald'' slower than a
spherically-symmetric Schwarzschild black hole.

(4) It has been widely accepted that the late-time tail of
gravitational collapse is {\it universal} in the sense that 
it is {\it independent} of the type of the massless field considered (e.g., scalar,
neutrino, electromagnetic, and gravitational). This belief was based
on {\it spherically} symmetric analyses. Our analysis, however,
turnover this point of view. In particular, the power indices $l+l_0+3$ at a fixed
radius which are found in this Letter are generically {\it different} from those
obtained in the scalar field toy-model \cite{Hod1,BarOr} $l+|m|+p+3$ (where $p=0$ if
$l-|m|$ is even, and $p=1$ otherwise). 

We have shown that different types of fields have
{\it different} decaying-rates. This is a rather surprising
conclusion, which has been overlooked in the last three decades ! 
It should be stressed, therefore, that the results obtained from the 
scalar field {\it toy}-model \cite{Hod1,BarOr} are actually 
{\it not} applicable for the physically interesting case of 
higher-spin perturbations (i.e., gravitational and
electromagnetic fields).

(5) Our results should have important implications for the mass-inflation scenario and the stability of
Cauchy horizons (see e.g., \cite{PoIs,BrDrMo} and references
therein). In particular, the late-time tails found in this Letter
should be used as initial data for perturbations propagating inside
the (rotating) black hole.

\bigskip
\noindent
{\bf ACKNOWLEDGMENTS}
\bigskip

I thank Tsvi Piran for discussions. 
This research was supported by a grant from the Israel Science Foundation.


\begin{thebibliography}{99}

\bibitem{Whee}R. Ruffini and J. A. Wheeler, Physics Today {\bf 24}, 30
  (1971); C. W. Misner, K. S. Thorne and J. A. Wheeler, Gravitation
  (Freeman, San Francisco 1973).

\bibitem{Price} R. H. Price, Phys. Rev. D {\bf 5}, 2419 (1972).

\bibitem{Thorne} K. S. Thorne, p. 231 in Magic without magic: John Archibald
Wheeler Ed: J.Klauder (W.H. Freeman, San Francisco 1972).

\bibitem {Hod2} S. Hod, Phys. Rev. D. (to be published); e-print gr-qc/9902073.

\bibitem{Krivan1} W. Krivan, P. Laguna and P. Papadopoulos,
  Phys. Rev. D {\bf 54}, 4728 (1996).

\bibitem{Krivan2} W. Krivan, P. Laguna and P. Papadopoulos and N. Andersson,
  Phys. Rev. D {\bf 56}, 3395 (1997).

\bibitem{Ori} A. Ori, Gen. Rel. Grav. {\bf 29}, Number 7, 881 (1997).
 
\bibitem{Barack2} L. Barack, in Internal structure of black holes and
  spacetime singularities, Volume XIII of the Israel Physical Society,
  Edited by L. M. Burko and A. Ori (Institute of Physics, Bristol, 1997).

\bibitem{Hod1} S. Hod, Phys. Rev. D. (to be published); e-print gr-qc/9902072.

\bibitem{BarOr} L. Barack and A. Ori, e-print gr-qc/9902082, Phys. Rev. Lett. {\bf 82}, 4388 (1999).

\bibitem{Teukolsky1} S. A. Teukolsky, Phys. Rev. Lett. {\bf 29}, 
1114 (1972).

\bibitem{Teukolsky2} S. A. Teukolsky, Astrophys. J. {\bf 185}, 635 (1973).

\bibitem{Andersson} N. Andersson, Phys. Rev. D {\bf 55}, 468 (1997).

\bibitem{Abram} M. Abramowitz and I.A. Stegun, Handbook of mathematical
functions (Dover Publications, New York 1970).

\bibitem{Leaver} E. W. Leaver, Phys. Rev. D {\bf 34}, 384 (1986).

\bibitem{Hod3} S. Hod, Phys. Rev. D {\bf 58}, 104022 (1998).

\bibitem{Schiff} L. Schiff, Quantum Mechanics, 3rd edition 
(McGraw-Hill, New York, 1968).

\bibitem{CamMor} W. B. Campbell and T. Morgan, Physica {\bf 53}, 264
  (1971).

\bibitem{BarOr2} L. Barack and A. Ori have recently shown
  (gr-qc/9907085, gr-qc/9908005) that $C_0$ vanishes in the particular
  case $am=0$ with $s>0$. Thus, 
$b=1$ in this non-generic case, whereas $b=0$ in all other cases.

\bibitem{PoIs} E. Poisson and W. Israel, Phys. Rev. D {\bf 41}, 1796 (1990).

\bibitem{BrDrMo} P. R. Brady, S. Droz and S. M. Morsink, Phys. Rev. D {\bf 58}, 084034 (1998).

\end{thebibliography}
\end{document}